\documentstyle[11pt]{article}
\begin{document}

\title{\textbf{Interacting viscous ghost tachyon, K-essence and dilaton scalar field models of dark energy}}


\author{K. Karami$^{1}$\thanks{KKarami@uok.ac.ir} , K. Fahimi$^{2}$\\
$^{1}$\small{Department of Physics, University of Kurdistan,
Pasdaran St., Sanandaj, Iran}\\$^{2}$\small{Department of Physics,
Sanandaj Branch, Islamic Azad
University, Sanandaj, Iran}\\
}

\maketitle

\begin{abstract}
We study the correspondence between the interacting viscous ghost
dark energy model with the tachyon, K-essence and dilaton scalar
field models in the framework of Einstein gravity. We consider a
spatially non-flat FRW universe filled with interacting viscous
ghost dark energy and dark matter. We reconstruct both the dynamics
and potential of these scalar field models according to the
evolutionary behavior of the interacting viscous ghost dark energy
model, which can describe the accelerated expansion of the universe.
Our numerical results show that the interaction and viscosity have
opposite effects on the evolutionary properties of the ghost scalar
filed models.
\end{abstract}



\clearpage
\section{GDE scenario}
The GDE density is proportional to the Hubble parameter
\cite{Urban,Witten,Forbes,Cai,sheikh,Sheykhi1,Sheykhi2,Sheykhi3,Rozas-Fernandez,KKaramiNew}
\begin{equation}
\rho_{D}=\alpha H,\label{GDE}
\end{equation}
where $\alpha$ is a constant. Here we consider a spatially non-flat
FRW universe filled with GDE and DM. Within the framework of FRW
cosmology, the first Friedmann equation takes the form
\begin{equation}
H^2+\frac{k}{a^2}=\frac{1}{3M_p^2}~ (\rho_{D}+\rho_{m}),\label{eqfr}
\end{equation}
where $M_{p}=(8\pi G)^{-1/2}$ is the reduced Planck mass. Here
$k=0,1,-1$ represent a flat, closed and open FRW universe,
respectively. Also $\rho_{D}$ and $\rho_{m}$ are the energy
densities of GDE and DM, respectively.

Using the dimensionless energy densities defined as
\begin{equation}
\Omega_{m}=\frac{\rho_{m}}{\rho_{\rm
cr}}=\frac{\rho_{m}}{3M_p^2H^2},~~~~\Omega_{D}=\frac{\rho_{D}}{\rho_{\rm
cr}}=\frac{\rho_{D}}{3M_p^2H^2},~~~~\Omega_{k}=\frac{k}{a^2H^2},
\label{eqomega}
\end{equation}
the Friedmann equation (\ref{eqfr}) can be rewritten as
\begin{equation}
1+\Omega_{k}=\Omega_{D}+\Omega_{m}.\label{eq10}
\end{equation}
Substituting Eq. (\ref{GDE}) into $\rho_{D}=3M_{p}^2H^2\Omega_{D}$
yields
 \begin{equation}\label{OmegaL}
   \Omega_{D}=\frac{\alpha}{3M_{p}^{2}H}.
 \end{equation}
Using the above relation, the curvature energy density parameter can
be obtained as
 \begin{equation}\label{omegak}
   \Omega_{k}=\left(\frac{9M_{p}^{4}k}{\alpha^{2}}\right)\left(\frac{\Omega_{D}}{a}\right)^2
   =\left(\frac{\Omega_{k_0}}{\Omega_{D_0}^2}\right)\left(\frac{\Omega_{D}}{a}\right)^2,
 \end{equation}
where we take $a_0=1$ for the present value of the scale factor.

Here, we extend our study to the viscous model of GDE. In the
presence of viscosity, the effective pressure of DE takes the form
\begin{equation}
\tilde{p}_D=p_D-3H\xi,\label{preshur}
\end{equation}
where $\xi=\varepsilon H^{-1}\rho_D$ is the bulk viscosity
coefficient in which $\varepsilon$ is a constant parameter
\cite{Zimdahl}. A viscosity $\varepsilon>0$ will be able to drive
acceleration \cite{Zimdahl}.

We further assume the viscous GDE interact with DM
\cite{Bertolami8}. In the presence of interaction, the continuity
equations are
\begin{equation}
\dot{\rho}_{D}+3H(1+\omega_{D})\rho_{D}=9\epsilon H \rho_{D}
-Q,\label{eqpol}
\end{equation}
\begin{equation}
\dot{\rho}_{m}+3H\rho_{m}=Q,\label{eqpol2}
\end{equation}
where $\omega_{D}=p_{D}/\rho_{D}$ is the equation of state (EoS)
parameter of the interacting viscous GDE and $Q$ stands for the
interaction term. Following \cite{Pavon}, we shall assume
$Q=3b^2H(\rho_{m}+\rho_{D})$ with the coupling constant $b^2$.

\begin{figure}
\includegraphics{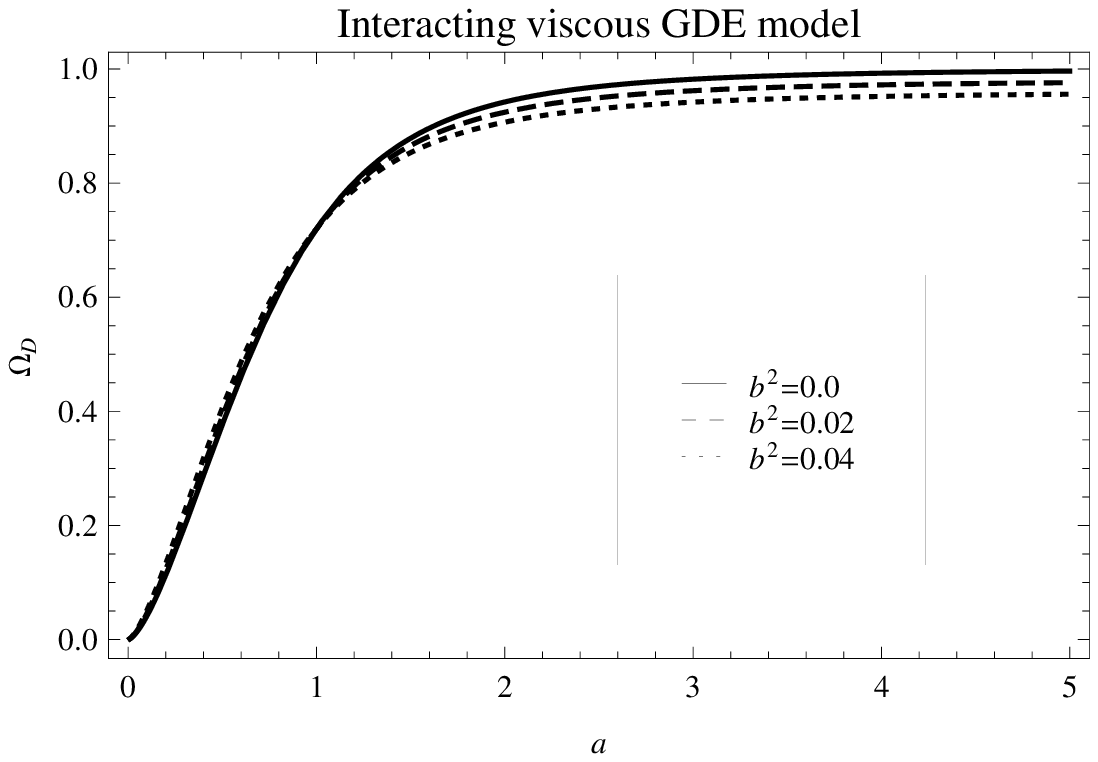}
      \vspace{4.7cm}
\caption[]{The evolution of the GDE density parameter, Eq.
(\ref{diffOmega}), for different coupling constants $b^2$. Auxiliary
parameters are $\Omega_{D_0}=0.72$ and $\Omega_{k_0}=0.01$.}
         \label{OmegaD}
   \end{figure}

Taking time derivative of Eq. (\ref{GDE}) and using Eqs.
(\ref{eqfr}), (\ref{eq10}), (\ref{OmegaL}) and (\ref{eqpol2}) gives
\begin{equation}
\frac{\dot{\rho}_{D}}{\rho_D}=3H\left[\frac{\Omega_D-1-\frac{\Omega_k}{3}+b^2(1+\Omega_k)}{2-\Omega_D}\right].\label{rhodot}
\end{equation}
Taking time derivative of Eq. (\ref{OmegaL}) and using (\ref{GDE})
and (\ref{rhodot}) one can obtain the evolution of the interacting
viscous GDE density parameter as
\begin{equation}\label{diffOmega}
\frac{{\rm d}\Omega_{D}}{{\rm d}\ln
a}=\left(\frac{3\Omega_D}{\Omega_D-2}\right)\left[\Omega_D-1-\frac{\Omega_k}{3}+b^2(1+\Omega_k)\right],
\end{equation}
which is same as that obtained for the interacting GDE in non-flat
universe in the absence of viscosity \cite{Sheykhi1}. It is
interesting to note that the viscosity constant $\epsilon$ does not
affect the evolution of the GDE density parameter (\ref{diffOmega}).
Substituting Eq. (\ref{omegak}) into (\ref{diffOmega}) yields a
differential equation for $\Omega_{D}(a)$ which can be solved
numerically with a suitable initial condition like
$\Omega_{D_0}=0.72$. The numerical results obtained for
$\Omega_{D}(a)$ are displayed in Fig. \ref{OmegaD} for different
coupling constant $b^2$. Figure shows that: i) for a given $b^2$,
$\Omega_D$ increases when the scale factor increases. ii) At early
and late times, $\Omega_D$ increases and decreases with increasing
$b^2$, respectively.

Substituting Eq. (\ref{rhodot}) into (\ref{eqpol}) gives the EoS
parameter of the interacting viscous GDE model as
\begin{equation}
\omega_{D}=\frac{1-\frac{\Omega_k}{3}+2b^2\Big(\frac{1+\Omega_k}{\Omega_D}\Big)}{\Omega_D-2}+3\epsilon,\label{wGDE}
\end{equation}
which shows that in the absence of interaction and viscous terms,
i.e. $b^2 =\epsilon= 0$, at early ($\Omega_{D}\rightarrow 0$) and
late ($\Omega_{D}\rightarrow 1$) times $\omega_{D}$ goes $-1/2$ and
$-1$, respectively, and cannot cross the phantom divide line
\cite{Cai}. For the present time ($a_0=1$), taking
$\Omega_{D_0}=0.72$ and $\Omega_{k_0}=0.01$ \cite{Bennett} Eq.
(\ref{wGDE}) gives
\begin{equation}
\omega_{D_0}=-0.78-2.19b^2+3\epsilon,\label{w1}
\end{equation}
which clears that for $\epsilon=0$ the phantom EoS parameter
($\omega_{D_0}<-1$) can be obtained provided $b^2>0.1$. This value
for $b^2$ is consistent with the recent observations in which we
have that $b^2$ could be as large as 0.2 \cite{Wang12}. Also the
phantom divide crossing is in accordance with the observations
\cite{Komatsu}.

\begin{figure}
\includegraphics{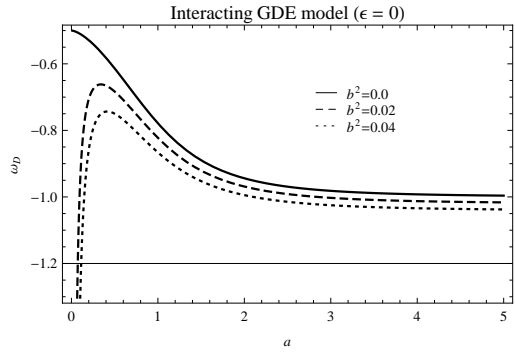}
      \vspace{4.7cm}
\caption[]{The evolution of the EoS parameter of GDE, Eq.
(\ref{wGDE}), for different coupling constants $b^2$ with
$\epsilon=0$. Auxiliary parameters as in Fig. \ref{OmegaD}.}
         \label{wD-b2}
   \end{figure}

\begin{figure}
\includegraphics{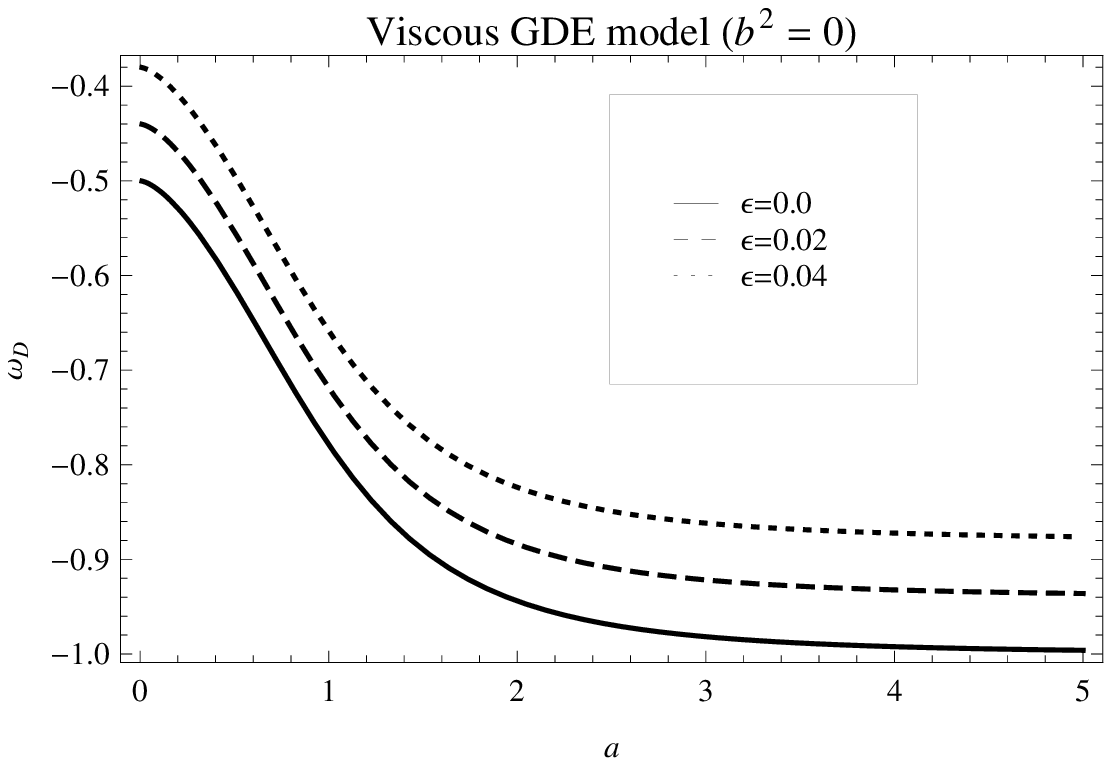}
      \vspace{4.7cm}
\caption[]{Same as Fig. \ref{wD-b2} for different viscosity
constants $\epsilon$ with $b^2=0$. Auxiliary parameters as in Fig.
\ref{OmegaD}.} \label{wD-e}
\end{figure}

The evolution of the EoS parameter (\ref{wGDE}) for different $b^2$
and $\epsilon$ is plotted in Figs. \ref{wD-b2} and \ref{wD-e},
respectively. Figure \ref{wD-b2} shows that: i) for $b^2=0$,
$\omega_D$ decreases from $-0.5$ at early times and approaches to
$-1$ at late times. ii) For $b^2\neq 0$, $\omega_D$ increases at
early times and decreases at late times. The results of $\omega_D$
in the absence of viscosity ($\epsilon=0$) are in agreement with
those obtained by \cite{Sheykhi1}. Figure \ref{wD-e} clears that: i)
for a given $\epsilon$, $\omega_D$ decreases with increasing the
scale factor. ii) For a given scale factor, $\omega_D$ increases
when $\epsilon$ increases.

\section{Ghost tachyon model}

The tachyon field is another approach for explaining DE. The tachyon
energy density and pressure are \cite{Sen}
\begin{equation}
\rho_{T}=\frac{V(\phi)}{\sqrt{1-\dot{\phi}^{2}}},\label{rhot}
\end{equation}
\begin{equation}
p_{T}=-V(\phi)\sqrt{1-\dot{\phi}^{2}}.
\end{equation}
The tachyon EoS parameter yields
\begin{equation}
\omega_{T}=\frac{p_{T}}{\rho_{T}}=\dot{\phi}^{2}-1.\label{wt}
\end{equation}

To reconstruct the tachyon filed via the interacting viscous GDE,
equating (\ref{wGDE}) with (\ref{wt}), i.e. $\omega_D=\omega_T$,
gives
\begin{equation}
\frac{1-\frac{\Omega_k}{3}+2b^2\Big(\frac{1+\Omega_k}{\Omega_D}\Big)}{\Omega_D-2}+3\epsilon=\dot{\phi}^{2}-1\label{aget1}.
\end{equation}
Also equating Eq. (\ref{GDE}) with (\ref{rhot}), i.e.
$\rho_D=\rho_T$, gets
\begin{equation}
\alpha H=\frac{V(\phi)}{\sqrt{1-\dot{\phi}^{2}}}\label{aget}.
\end{equation}
From Eqs. (\ref{aget1}) and (\ref{aget}), the kinetic energy and
potential of the tachyon field can be obtained as follows
\begin{equation}
\dot{\phi}^{2}=\frac{\Omega_D-1-\frac{\Omega_k}{3}
+2b^2\Big(\frac{1+\Omega_k}{\Omega_D}\Big)}{\Omega_D-2}+3\epsilon\label{phi2},
\end{equation}
\begin{equation}
V(\phi)=\frac{\alpha^2}{3M_p^2\Omega_D}
\left[\frac{1-\frac{\Omega_k}{3}+2b^2\Big(\frac{1+\Omega_k}{\Omega_D}\Big)}{2-\Omega_D}
-3\epsilon\right]^{1/2}\label{vphi1}.
\end{equation}
Note that Eqs. (\ref{phi2}) and (\ref{vphi1}) for the flat case,
i.e. $\Omega_k=0$, and in the absence of viscosity ($\epsilon=0$)
reduce to the results obtained by \cite{Sheykhi3}.

\begin{figure}
\includegraphics{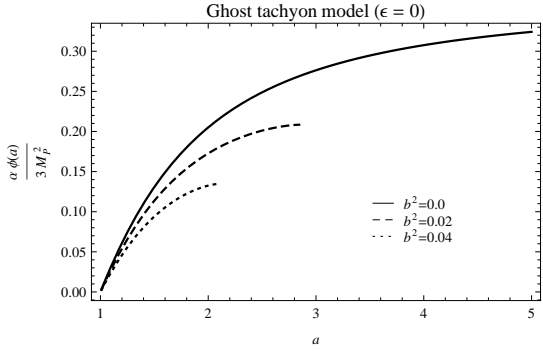}
      \vspace{4.7cm}
\caption[]{The evolution of the ghost tachyon scalar filed, Eq.
(\ref{phi3}), for different coupling constants $b^2$ with
$\epsilon=0$. Auxiliary parameters are $\Omega_{D_0}=0.72$,
$\Omega_{k_0}=0.01$ and $\phi(1)=0$.}
         \label{Phi-Tachyon-b2}
   \end{figure}

\begin{figure}
\includegraphics{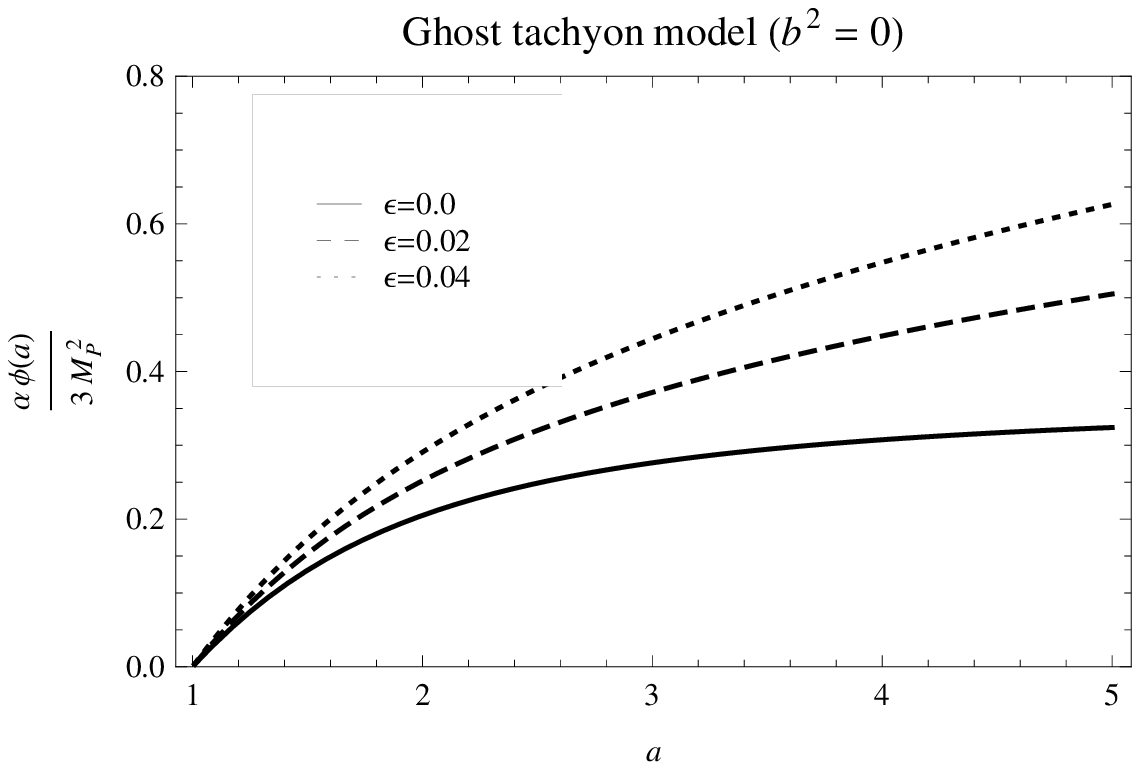}
      \vspace{4.7cm}
\caption[]{Same as Fig. \ref{Phi-Tachyon-b2} for different viscosity
constants $\epsilon$ with $b^2=0$. Auxiliary parameters as in Fig.
\ref{Phi-Tachyon-b2}.}
         \label{Phi-Tachyon-e}
   \end{figure}

\begin{figure}
\includegraphics{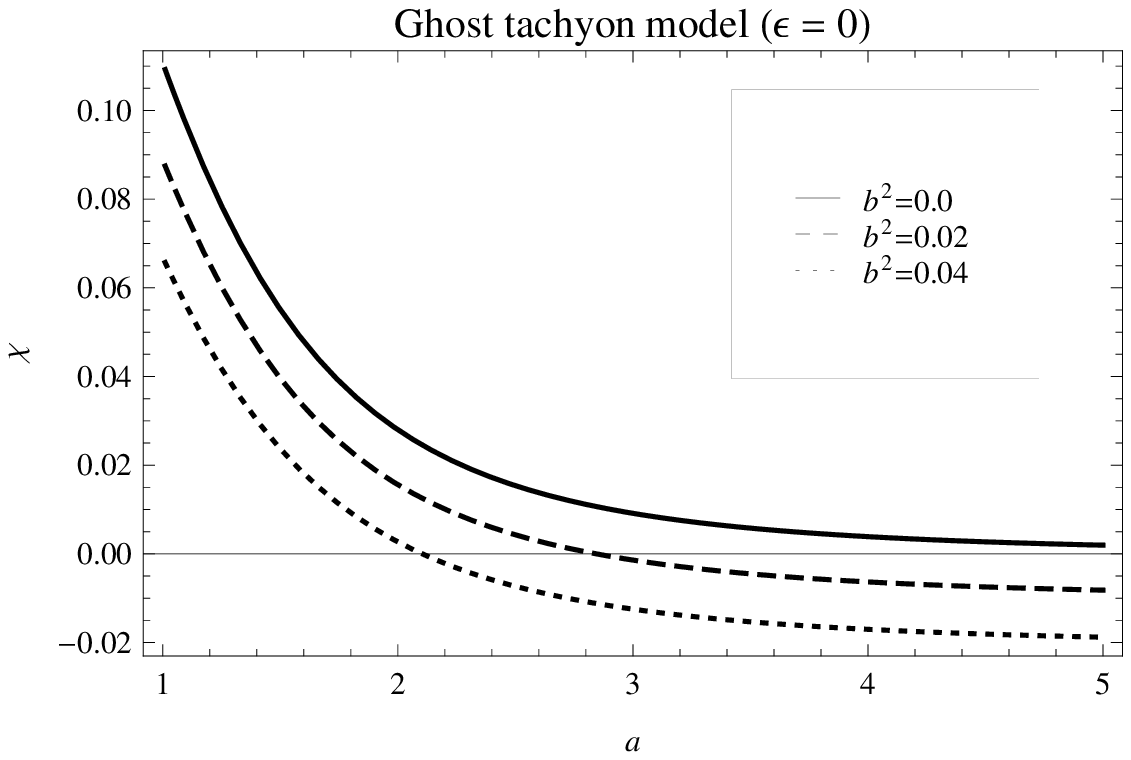}
      \vspace{4.7cm}
\caption[]{The evolution of the ghost tachyon kinetic energy
$\chi=\dot{\phi}^2/2$, Eq. (\ref{phi2}), for different coupling
constants $b^2$ with $\epsilon=0$. Auxiliary parameters as in Fig.
\ref{Phi-Tachyon-b2}.}
         \label{Phidot2-Tachyon-b2}
   \end{figure}

\begin{figure}
\includegraphics{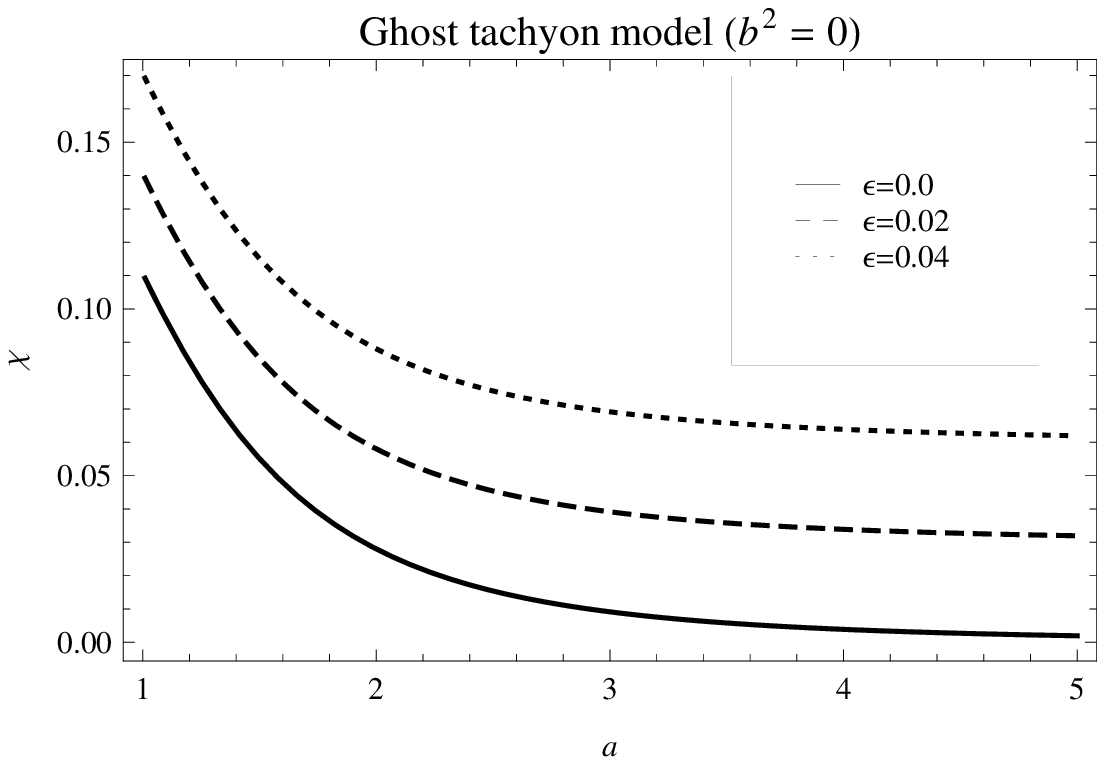}
      \vspace{4.7cm}
\caption[]{Same as Fig. \ref{Phidot2-Tachyon-b2} for different
viscosity constants $\epsilon$ with $b^2=0$. Auxiliary parameters as
in Fig. \ref{Phi-Tachyon-b2}.}
         \label{Phidot2-Tachyon-e}
   \end{figure}

From Eq. (\ref{phi2}) and using (\ref{OmegaL}), one can get the
evolutionary form of the ghost tachyon scalar field as
\begin{equation}
\phi(a)-\phi(1)=\frac{3M_p^2}{\alpha}\int_{1}^{a}\Omega_D
\left[\frac{\Omega_D-1-\frac{\Omega_k}{3}+2b^2\Big(\frac{1+\Omega_k}{\Omega_D}\Big)}{\Omega_D-2}
+3\epsilon\right]^{1/2}\frac{{\rm d}a}{a}\label{phi3},
\end{equation}
where we take $a_0=1$ for the present time. The evolution of the
ghost tachyon scalar filed, Eq. (\ref{phi3}), for different values
of $b^2$ and $\epsilon$ is plotted in Figs. \ref{Phi-Tachyon-b2} and
\ref{Phi-Tachyon-e}, respectively. Figures clear that: i) for a
given $b^2$ or $\epsilon$, $\phi(a)$ increases with increasing the
scale factor. ii) For a given scale factor, $\phi(a)$ decreases and
increases with increasing $b^2$ and $\epsilon$, respectively. Note
that Fig. \ref{Phi-Tachyon-b2} shows only the real scalar field,
i.e. $\dot{\phi}^2>0$. Indeed, for $b^2=0$, $0.02$ and $0.04$ the
scalar field $\phi$ becomes pure imaginary ($\dot{\phi}^2<0$) at
$a>43.5$, $2.8$ and $2.1$, respectively, and it does not show itself
in Fig. \ref{Phi-Tachyon-b2}. To investigate this problem in ample
detail, the evolution of the ghost tachyon kinetic energy
$\chi=\dot{\phi}^2/2$, Eq. (\ref{phi2}), for different values of
$b^2$ and $\epsilon$ is plotted in Figs. \ref{Phidot2-Tachyon-b2}
and \ref{Phidot2-Tachyon-e}, respectively. Figure
\ref{Phidot2-Tachyon-b2} confirms that for $b^2=0$, $0.02$ and
$0.04$ the kinetic energy becomes negative ($\chi<0$) at $a>43.5$,
$2.8$ and $2.1$, respectively. Figures \ref{Phidot2-Tachyon-b2} and
\ref{Phidot2-Tachyon-e} show that: i) for a given $b^2$ or
$\epsilon$, the kinetic energy $\chi$ decreases when the scale
factor increases. ii) For a given scale factor, the kinetic energy
decreases and increases with increasing $b^2$ and $\epsilon$,
respectively.

It is worth to note that from Eq. (\ref{rhot}) due to having a real
tachyon energy density we need to have $\dot{\phi}^2<1$ which is in
accordance with Figs. \ref{Phidot2-Tachyon-b2} and
\ref{Phidot2-Tachyon-e}. Moreover, from Eq. (\ref{wt}) for
$\dot{\phi}^2<0$ and $0<\dot{\phi}^2<1$ we have $\omega_T<-1$ and
$-1<\omega_T<0$, respectively, corresponding to the phantom
\cite{Caldwell} and quintessence \cite{Ratra} DE, respectively. In
the absence of interaction ($b^2=0$), the kinetic energy of the
ghost tachyon scalar field is always positive (see Fig.
\ref{Phidot2-Tachyon-e}) and behaves like quintessence DE with
$\omega_T=\omega_D>-1$ (see Fig. \ref{wD-e}).

\begin{figure}
\includegraphics{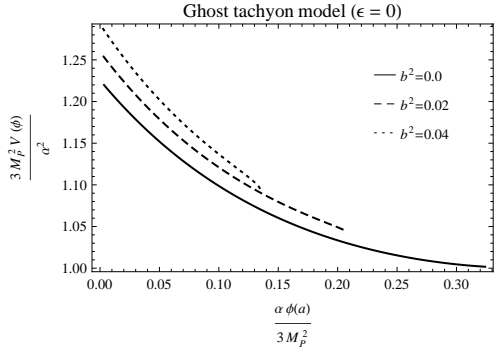}
      \vspace{4.7cm}
\caption[]{The ghost tachyon potential, Eq. (\ref{vphi1}), versus
the scalar field $\phi$ for different coupling constants $b^2$ with
$\epsilon=0$. Auxiliary parameters as in Fig. \ref{Phi-Tachyon-b2}.}
         \label{V-Phi-Tachyon-b2}
   \end{figure}

\begin{figure}
\includegraphics{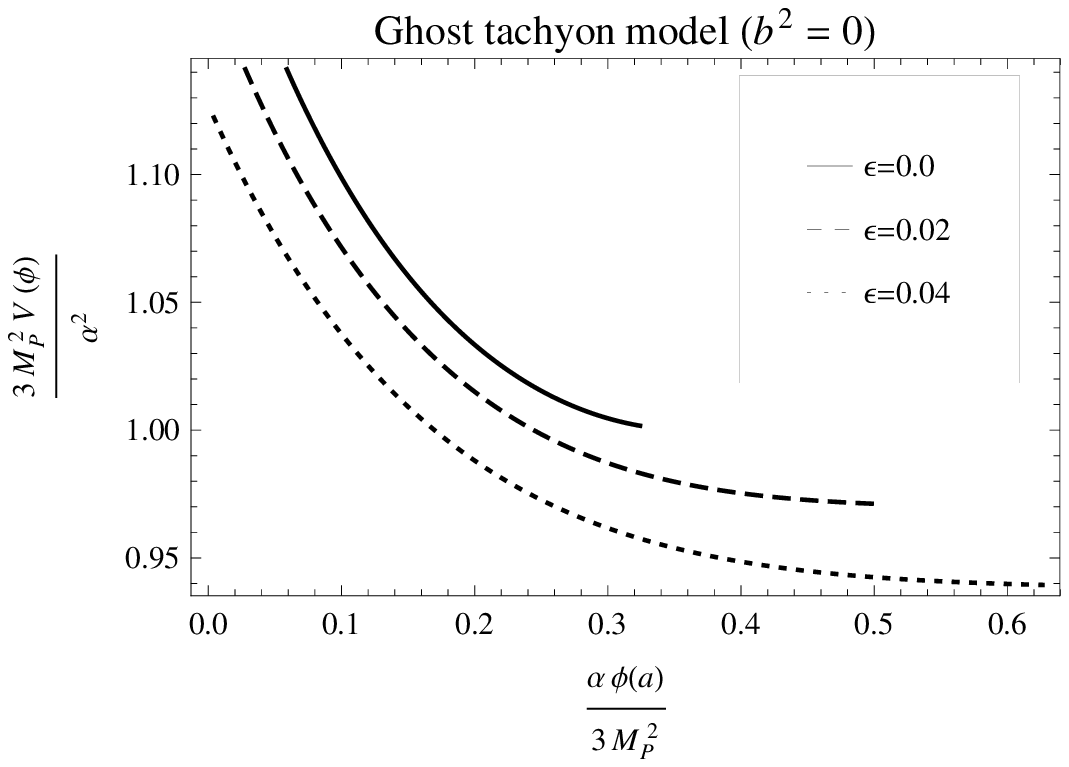}
      \vspace{4.7cm}
\caption[]{Same as Fig. \ref{V-Phi-Tachyon-b2} for different
viscosity constants $\epsilon$ with $b^2=0$. Auxiliary parameters as
in Fig. \ref{Phi-Tachyon-b2}.}
         \label{V-Phi-Tachyon-e}
   \end{figure}

The ghost tachyon potential, Eq. (\ref{vphi1}), versus the scalar
field (\ref{phi3}) for different $b^2$ and $\epsilon$ is plotted in
Figs. \ref{V-Phi-Tachyon-b2} and \ref{V-Phi-Tachyon-e},
respectively. Figures illustrate that: i) for a given $b^2$ or
$\epsilon$, $V(\phi)$ decreases with increasing $\phi$. This
behavior is in agreement
 with the scaling solution $V(\phi)\propto
\phi^{-2}$ obtained for the tachyon filed corresponding to the power
law expansion \cite{Copeland1}. ii) For a given scalar field,
$V(\phi)$ increases and decreases with increasing $b^2$ and
$\epsilon$, respectively.
\section{Ghost K-essence model}

\begin{figure}
\includegraphics{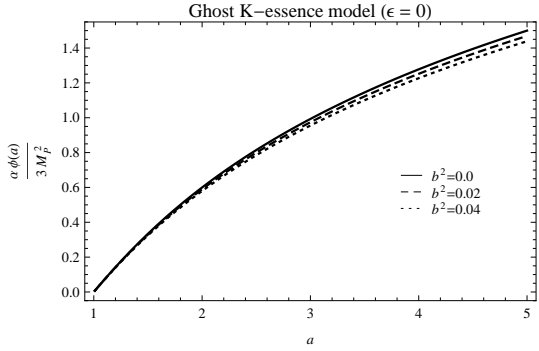}
      \vspace{4.7cm}
\caption[]{The evolution of the ghost K-essence scalar field, Eq.
(\ref{phik}), for different coupling constants $b^2$ with
$\epsilon=0$. Auxiliary parameters as in Fig. \ref{Phi-Tachyon-b2}.}
         \label{Phi-K-essence-b2}
   \end{figure}

\begin{figure}
\includegraphics{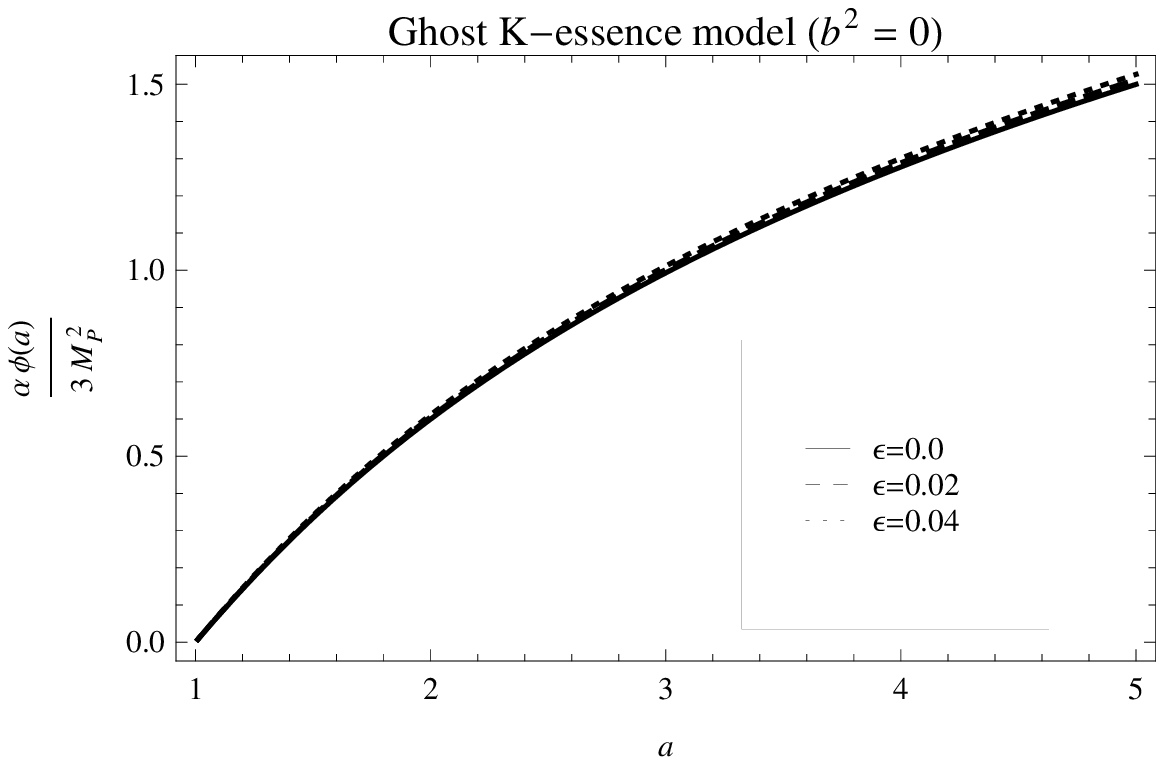}
      \vspace{4.7cm}
\caption[]{Same as Fig. \ref{Phi-K-essence-b2} for different
viscosity constants $\epsilon$ with $b^2=0$. Auxiliary parameters as
in Fig. \ref{Phi-Tachyon-b2}.}
         \label{Phi-K-essence-e}
   \end{figure}

The K-essence scalar field model of DE is given by the action
\cite{Chiba, Picon3}
\begin{equation}
S=\int {\rm d} ^{4}x\sqrt{-{\rm g}}~p(\phi,\chi),\label{action}
\end{equation}
where $p(\phi,\chi)$ is the Lagrangian density given by
\begin{equation}
p(\phi,\chi)=f(\phi)(-\chi+\chi^{2}),\label{pk}
\end{equation}
and the K-essence energy density is
\begin{equation}
\rho(\phi,\chi)=f(\phi)(-\chi+3\chi^{2}).\label{rhok}
\end{equation}
The K-essence EoS parameter takes the form
\begin{equation}
\omega_{K}=\frac{p(\phi,\chi)}{\rho(\phi,\chi)}=\frac{\chi-1}{3\chi-1}.\label{wk}
\end{equation}
Equating (\ref{wk}) with (\ref{wGDE}), $\omega_{K}=\omega_{D}$, we
get
\begin{equation}
\chi=\frac{3-\frac{\Omega_k}{3}+2b^2\Big(\frac{1+\Omega_k}{\Omega_D}\Big)-\Omega_D+3\epsilon(\Omega_D-2)}
{5-\Omega_k+6b^2\Big(\frac{1+\Omega_k}{\Omega_D}\Big)
-\Omega_D+9\epsilon(\Omega_D-2)}\label{chi}.
\end{equation}
Using Eq. (\ref{chi}) and $\dot{\phi}^2=2\chi$, we obtain the ghost
K-essence scalar field as
\begin{equation}
\phi(a)-\phi(1)=\frac{3M_p^2}{\alpha}\int_1^a
\Omega_{D}\left[\frac{6-\frac{2\Omega_k}{3}+4b^2\Big(\frac{1+\Omega_k}{\Omega_D}\Big)
-2\Omega_D+6\epsilon(\Omega_D-2)}{5-\Omega_k+6b^2\Big(\frac{1+\Omega_k}{\Omega_D}\Big)
-\Omega_D+9\epsilon(\Omega_D-2)}\right]^{1/2}\frac{{\rm d}a}{a}
\label{phik},
\end{equation}
which its evolution for different $b^2$ and $\epsilon$ is displayed
in Figs. \ref{Phi-K-essence-b2} and \ref{Phi-K-essence-e},
respectively. Figures present that: i) for a given $b^2$ or
$\epsilon$, $\phi(a)$ increases with increasing the scale factor.
ii) For a given scale factor, $\phi(a)$ decreases and increases with
increasing $b^2$ and $\epsilon$, respectively.

\begin{figure}
\includegraphics{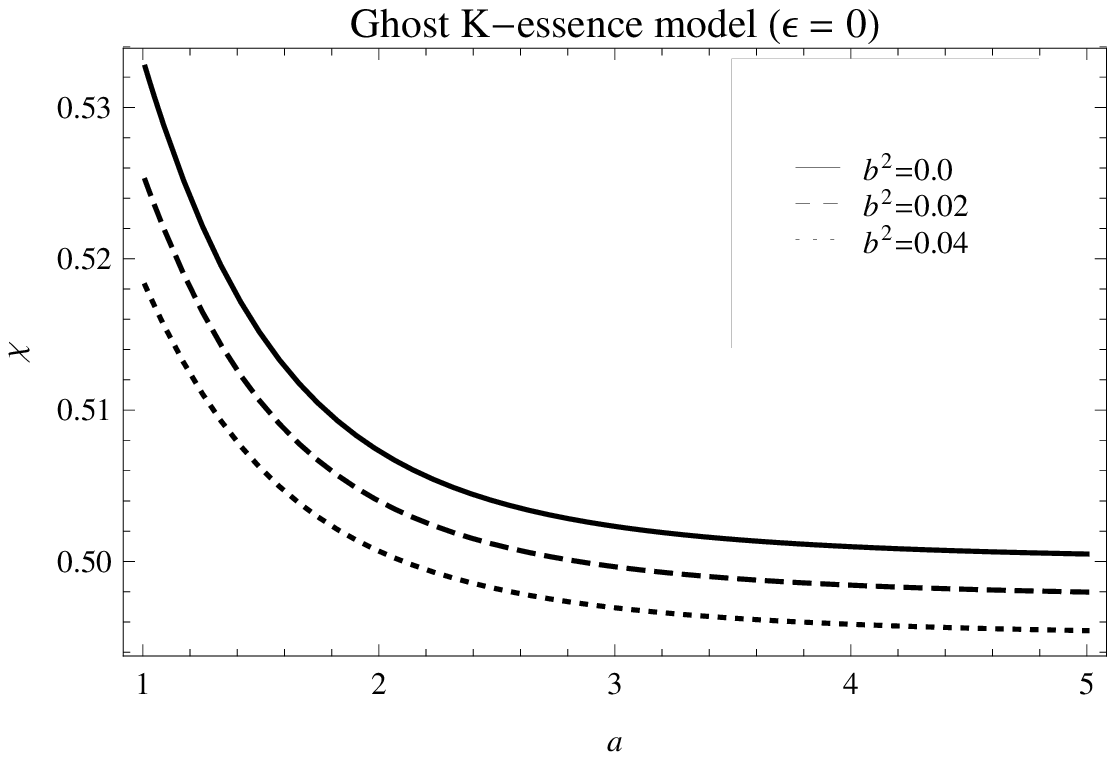}
      \vspace{4.7cm}
\caption[]{The evolution of the ghost K-essence kinetic energy
$\chi=\dot{\phi}^2/2$, Eq. (\ref{chi}), for different coupling
constants $b^2$ with $\epsilon=0$. Auxiliary parameters as in Fig.
\ref{Phi-Tachyon-b2}.}
         \label{Phidot2-K-essence-b2}
   \end{figure}

\begin{figure}
\includegraphics{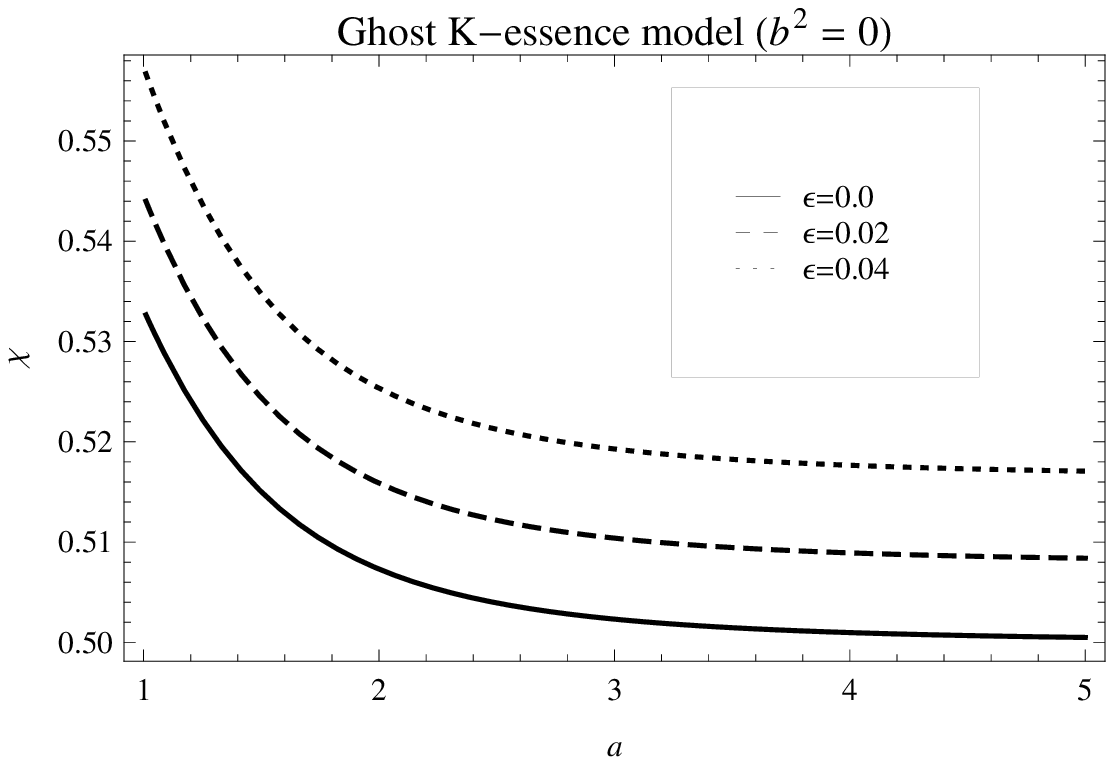}
      \vspace{4.7cm}
\caption[]{Same as Fig. \ref{Phidot2-K-essence-b2} for different
viscosity constants $\epsilon$ with $b^2=0$. Auxiliary parameters as
in Fig. \ref{Phi-Tachyon-b2}.}
         \label{Phidot2-K-essence-e}
   \end{figure}

The evolution of the ghost K-essence kinetic energy, Eq.
(\ref{chi}), for different values of $b^2$ and $\epsilon$ is plotted
in Figs. \ref{Phidot2-K-essence-b2} and \ref{Phidot2-K-essence-e},
respectively. Figures clarify that: i) for a given $b^2$ or
$\epsilon$, the ghost K-essence kinetic energy like the tachyon
filed decreases when the scale factor increases. ii) For a given
scale factor, the kinetic energy of the ghost K-essence filed like
the tachyon model decreases and increases with increasing $b^2$ and
$\epsilon$, respectively. If we compare Fig.
\ref{Phidot2-K-essence-b2} with \ref{Phidot2-Tachyon-b2} we see that
the kinetic energy of the ghost K-essence model in contrast with the
ghost tachyon field is always positive. Note that the result of Fig.
\ref{Phidot2-K-essence-b2} is in contrast with that obtained by
\cite{Rozas-Fernandez} who showed that for a given $b^2$, the
kinetic energy of the ghost K-essence filed increases with
increasing the scale factor. This difference may come back to this
fact that the K-essence filed selected by \cite{Rozas-Fernandez} is
a purely kinetic model in which the action (\ref{action}) is
independent of $\phi$. This yields the energy density and pressure
of a purely kinetic K-essence which are different from those
considered in Eqs. (\ref{pk}) and (\ref{rhok}).

\section{Ghost dilaton model}

The pressure and energy density of the dilaton scalar field model
are given by \cite{Gasperini}
\begin{equation}
p_{D}=-\chi+ce^{\lambda\phi}\chi^{2},
\end{equation}
\begin{equation}
\rho_{D}=-\chi+3ce^{\lambda\phi}\chi^{2},\label{rhod}
\end{equation}
where $c$ and $\lambda$ are constants and $\chi=\dot{\phi}^2/2$. The
dilaton EoS parameter takes the form
\begin{equation}
\omega_{D}=\frac{p_D}{\rho_D}=\frac{ce^{\lambda\phi}\chi-1}{3ce^{\lambda\phi}\chi-1}.\label{wd}
\end{equation}
Equating (\ref{wd}) with (\ref{wGDE}) gives the solution
\begin{equation}
ce^{\lambda\phi}\chi=\frac{3-\frac{\Omega_k}{3}+2b^2\Big(\frac{1+\Omega_k}{\Omega_D}\Big)
-\Omega_D+3\epsilon(\Omega_D-2)}{5-\Omega_k+6b^2
\Big(\frac{1+\Omega_k}{\Omega_D}\Big)-\Omega_D+9\epsilon(\Omega_D-2)},\label{chidilaton}
\end{equation}
then with the help of $\chi=\dot{\phi}^2/2$, we obtain
\begin{equation}
e^{\frac{\lambda\phi}{2}}\dot{\phi}=\sqrt{\frac{2}{c}}
\left[\frac{3-\frac{\Omega_k}{3}+2b^2\Big(\frac{1+\Omega_k}{\Omega_D}\Big)
-\Omega_D+3\epsilon(\Omega_D-2)}{5-\Omega_k+6b^2\Big(\frac{1+\Omega_k}{\Omega_D}\Big)
-\Omega_D+9\epsilon(\Omega_D-2)}\right]^{1/2} \label{phiD1}.
\end{equation}
Finally we obtain
\begin{equation}
\phi(a)=\frac{2}{\lambda}\ln\left\{e^\frac{\lambda\phi(1)}{2}+\frac{3M_p^2\lambda}{2\alpha\sqrt{c}}\int_1^a\Omega_D
\left[\frac{6-\frac{2\Omega_k}{3}+4b^2\Big(\frac{1+\Omega_k}{\Omega_D}\Big)
-2\Omega_D+6\epsilon(\Omega_D-2)}{5-\Omega_k+6b^2\Big(\frac{1+\Omega_k}{\Omega_D}\Big)
-\Omega_D+9\epsilon(\Omega_D-2)}\right]^{1/2}\frac{{\rm
d}a}{a}\right\}\label{phiD}.
\end{equation}

\begin{figure}
\includegraphics{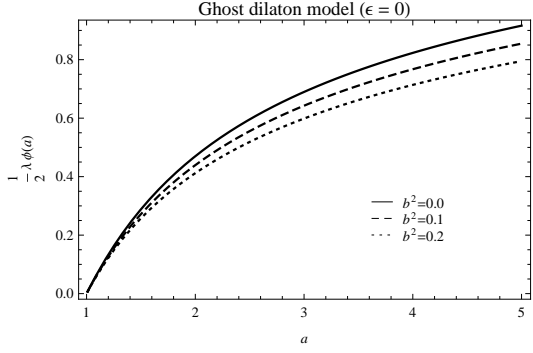}
      \vspace{4.7cm}
\caption[]{The evolution of the ghost dilaton scalar field, Eq.
(\ref{phiD}), for different coupling constants $b^2$ with
$\epsilon=0$. Auxiliary parameters are $\Omega_{D_0}=0.72$,
$\Omega_{k_0}=0.01$, $\phi(1)=0$ and
$\frac{3M_p^2\lambda}{2\alpha\sqrt{c}}=1$. }
         \label{Phi-Dilaton-b2}
   \end{figure}

\begin{figure}
\includegraphics{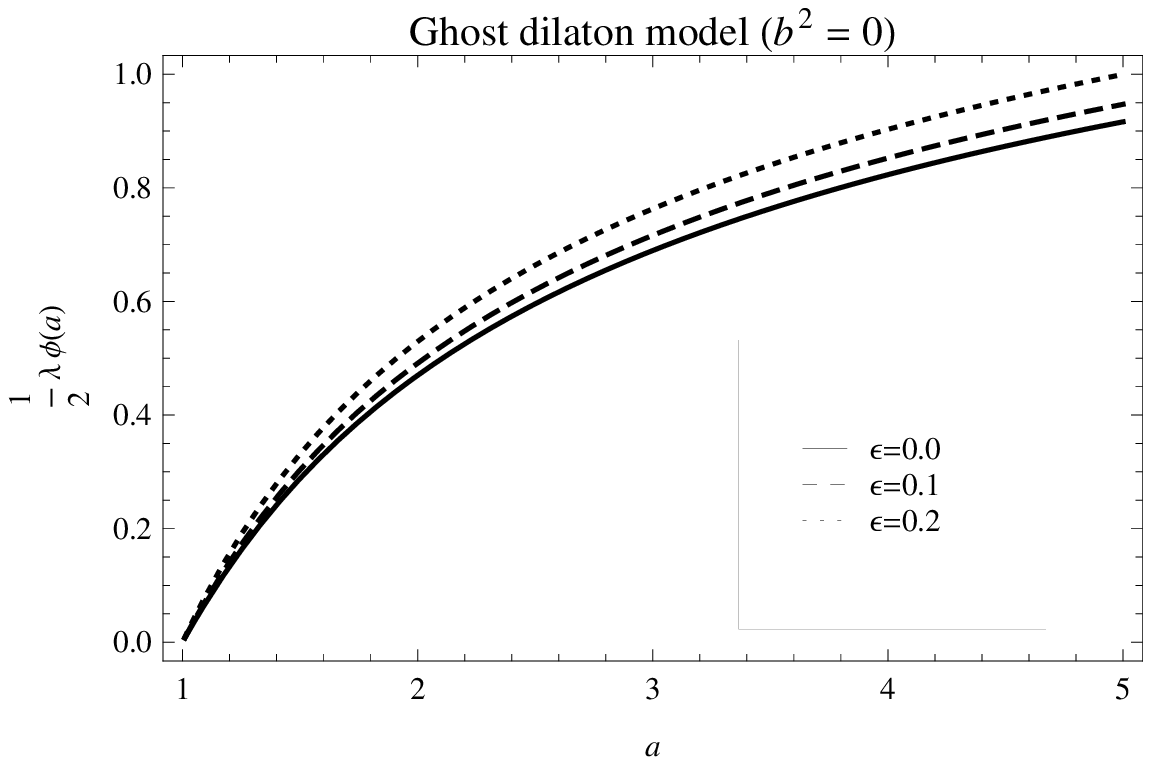}
      \vspace{4.7cm}
\caption[]{Same as Fig. \ref{Phi-Dilaton-b2} for different viscosity
constants $\epsilon$ with $b^2=0$. Auxiliary parameters as in Fig.
\ref{Phi-Dilaton-b2}.}
         \label{Phi-Dilaton-e}
   \end{figure}

The evolution of the ghost dilaton scalar field (\ref{phiD}) for
different $b^2$ and $\epsilon$ is displayed in Figs.
\ref{Phi-Dilaton-b2} and \ref{Phi-Dilaton-e}, respectively. Figures
present that: i) for a given $b^2$ or $\epsilon$, $\phi(a)$
increases with increasing the scale factor. ii) For a given scale
factor, $\phi(a)$ decreases and increases with increasing $b^2$ and
$\epsilon$, respectively.

\begin{figure}
\includegraphics{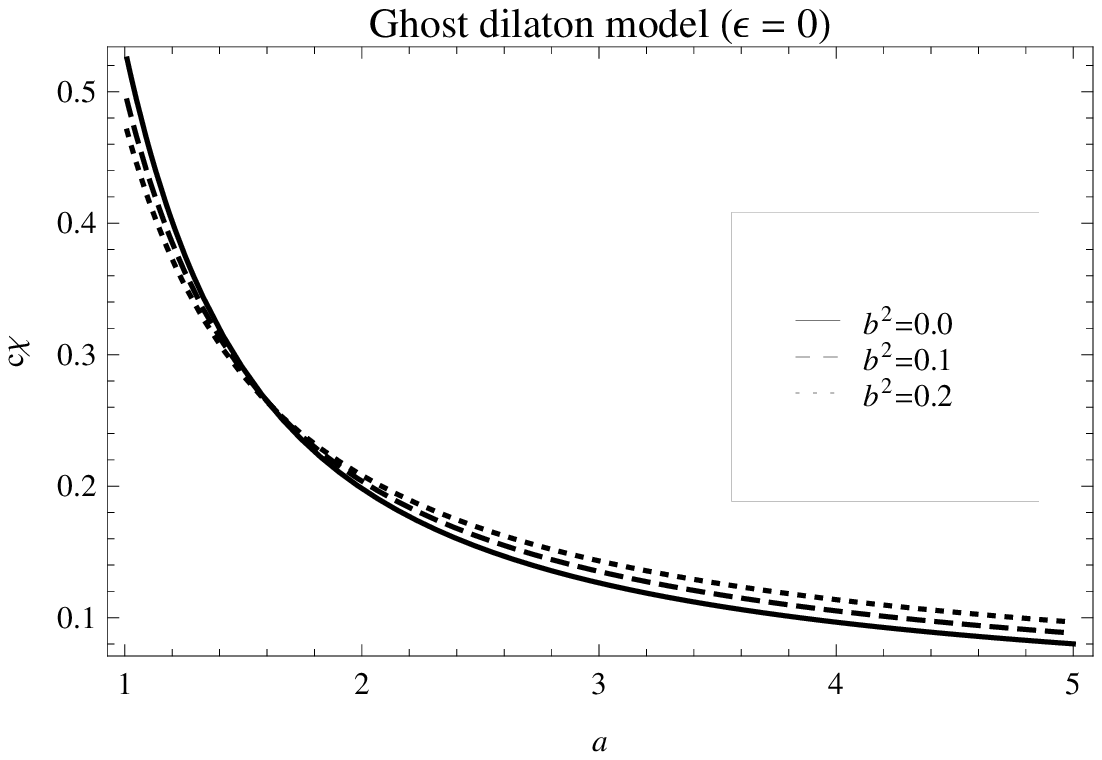}
      \vspace{4.7cm}
\caption[]{The evolution of the ghost dilaton kinetic energy
$\chi=\dot{\phi}^2/2$, Eq. (\ref{chidilaton}), for different
coupling constants $b^2$ with $\epsilon=0$. Auxiliary parameters as
in Fig. \ref{Phi-Dilaton-b2}.}
         \label{Phidot2-Dilaton-b2}
   \end{figure}

\begin{figure}
\includegraphics{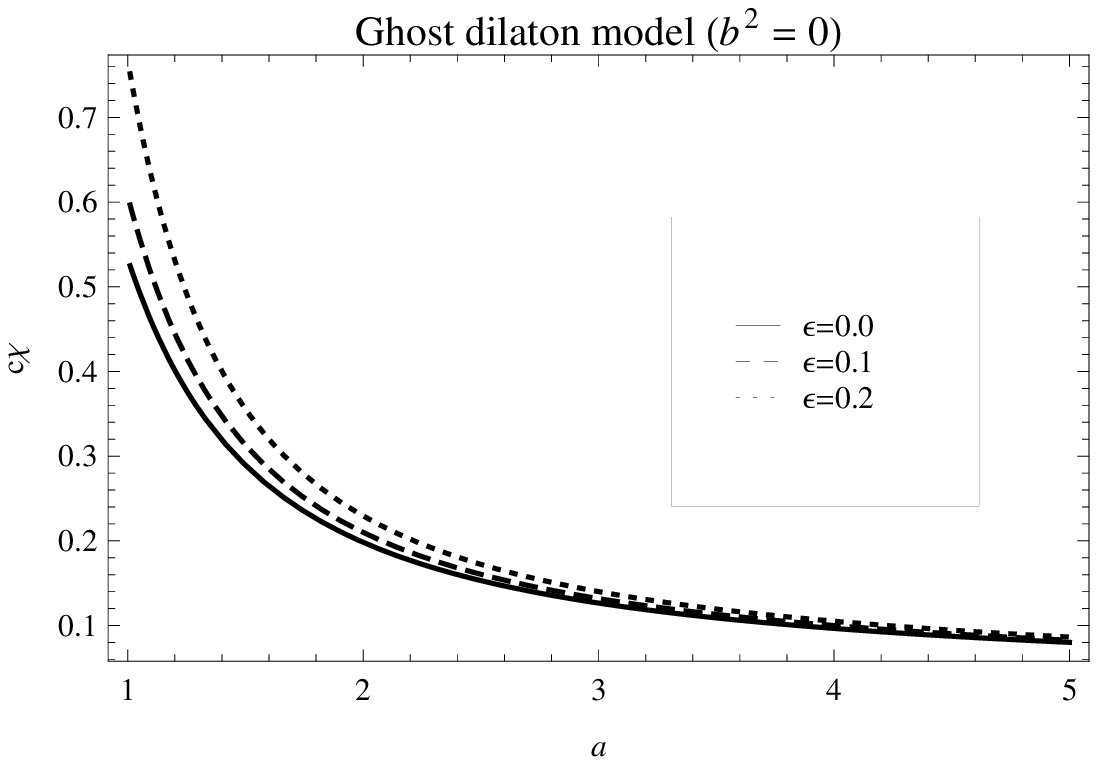}
      \vspace{4.7cm}
\caption[]{Same as Fig. \ref{Phidot2-Dilaton-b2} for different
viscosity constants $\epsilon$ with $b^2=0$. Auxiliary parameters as
in Fig. \ref{Phi-Dilaton-b2}.}
         \label{Phidot2-Dilaton-e}
   \end{figure}

With the help of Eq. (\ref{chidilaton}) we plot the evolution of the
ghost dilaton kinetic energy for different $b^2$ and $\epsilon$ in
Figs. \ref{Phidot2-Dilaton-b2} and \ref{Phidot2-Dilaton-e},
respectively. Figures show that for a given $b^2$ or $\epsilon$, the
kinetic energy of the ghost dilaton field like the tachyon and
K-essence models decreases with increasing the scale factor.

\section{Conclusions}
Here we investigated the interacting viscous GDE model in the
framework of standard FRW cosmology. For a spatially non-flat FRW
universe containing GDE and DM, we obtained the evolution of the
fractional energy density and EoS parameters of the interacting
viscous GDE model throughout history of the universe. Furthermore,
we reconstructed both the dynamics and potential of the tachyon,
K-essence and dilaton scalar filed models of DE according the
evolutionary behavior of the interacting viscous GDE model. Our
numerical results show that:

(i) The evolution of the interacting viscous GDE density parameter
$\Omega_D$ is independent of viscosity constant $\epsilon$. But for
a given coupling constant $b^2$, $\Omega_D$ increases with
increasing the scale factor. Also at early and late times,
$\Omega_D$ increases and decreases, respectively, with increasing
$b^2$.

(ii) The EoS parameter $\omega_D$ of the GDE model in the absence of
viscosity, can cross the phantom divide line ($\omega_D<-1$) at the
present provided $b^2>0.1$ which is compatible with the
observations. Also in the absence of viscosity for a given coupling
constant $b^2$, $\omega_D$ increases and decreases at early and late
times, respectively. Moreover, in the absence of interaction for a
given viscosity constant $\epsilon$, $\omega_D$ decreases when the
scale factor increases. For a given scale factor, $\omega_D$
increases with increasing $\epsilon$.

(iii) The ghost tachyon scalar filed for a given $b^2$ or
$\epsilon$, increases with increasing the scale factor. Also for a
given scale factor, it decreases and increases with increasing $b^2$
and $\epsilon$, respectively. For a given $b^2$ or $\epsilon$, the
ghost tachyon kinetic energy $\chi(a)$ and potential $V(\phi)$
decrease with increasing the scale factor and scalar filed,
respectively. For a given scale factor, $\chi(a)$ decreases and
increases with increasing $b^2$ and $\epsilon$, respectively. For a
given scalar field, $V(\phi)$ increases and decreases with
increasing $b^2$ and $\epsilon$, respectively.

(iv) The ghost K-essence scalar filed for a given $b^2$ or
$\epsilon$ increases with increasing the scale factor. But its
kinetic energy decreases. For a given scale factor, the ghost
K-essence scalar filed decreases and increases with increasing $b^2$
and $\epsilon$, respectively. This behavior also holds for the
kinetic energy of the ghost K-essence model.

(v) The ghost dilaton scalar filed and its corresponding kinetic
energy for a given $b^2$ or $\epsilon$ behave like the ghost
K-essence model.

All mentioned in above illustrate that the interaction and viscosity
have opposite effects on the dynamics of ghost tachyon, K-essence
and dilaton scalar field models of DE.

\clearpage
\clearpage
\clearpage
\clearpage
\clearpage
\clearpage
\clearpage
\clearpage
\end{document}